# Comment on "Interference in the Collective Electron Momentum in Double Photoionization of H$_2$"

G. Van Hooydonk, Faculty of Sciences, Ghent University, Ghent, Belgium

**Abstract.** We give supporting evidence for and elaborate further on the fact [see Kreidi et al, Phys. Rev. Lett. **100**, 133005 (2008)] that *ionic* states are essential for *covalent* dihydrogen H$_2$.

Kreidi et al [1] proved that interference in double photoionization of H$_2$ results *from a non-Heitler-London fraction of the H$_2$ ground state, where both electrons are at the same atomic center*, i.e. photoionization reveals *ionic* states for *covalent* H$_2$ [1]. We give supporting experimental evidence for [1] and report on bond theories of ionic type, justified by [1] but not fully elaborated therein.

(i) Whenever 2 electrons are at 1 nucleon [1], the HL bonding region in between the 2 nucleons is deprived of electron-density. These remarkable details on the internal mechanics of H$_2$ [1] are supported by Dunitz and Seiler [2], who found earlier that, for covalent bonds, the HL bonding region between the nuclei is *void of electron-density*. Hence, both [1] and [2] allow for an *ionic* view on H$_2$. At the time, results [2] faced fierce opposition from theorists [3]. With [1,2], such opposition must be moderated and a search for refined, if not alternative bond theories is justified [4].

(ii) Although Kreidi et al. [1] argue that non-HL *ionic* states [5] are indeed needed, they do not say how important these states really are. Knowing that HL-theory uses 0% of ionic states, they refer [1] to VB-theory [5] with a small % of ionic states but not to MO-theory with 50 % or to ionic theory with 100% of ionic contributions to the H$_2$ ground state. For instance, 19$^{th}$ century ionic models [4,6] applied to covalent H$_2$ give H$_2$=[H$^+$H$^-$+H$^-$H$^+$], where 2 electrons are at 1 nucleon, either at the right +½r$_0$ or at the left -½r$_0$ of the center of mass (or vice versa), as in Fig. 2 of [1]. It is obvious that any ionic contribution for covalent H$_2$, however large, can only make sense

a) if its 2 valence electrons were centered at one atomic center instead of the two [1]
b) if electrons are absent in the HL-bonding region [2]. For *a 100% ionic bond* however,
c) the driving force must be Coulomb attraction –e$^2$/r [4,6], where r is (close to) the internucleon separation. While qualitative constraints a and b are met with [1,2], even rather severe quantitative constraint c for a 100 % ionic bond is obeyed, since ionic bond energy D$_{ion}$=-e$^2$/r$_0$ rationalizes lower order spectroscopic constants *of diatomic ionic and covalent bonds* between all monovalent atoms in the periodic table, *including covalent bond* H$_2$ [4,7,8]. Despite this, ionic models give problems, although equilibrium value –e$^2$/r$_0$= 19,5 eV for H$_2$ is largely encompassed by photon energies varying from 130 to 240 eV used in [1].

QM bond theories use the Born-Oppenheimer approximation (BOA) [7], where bond stability depends on *internucleon repulsion* +e$^2$/r, which is *mutually exclusive* with *attraction* –e$^2$/r. Hence, only one of the two can decide on bond stability [10]. Born's BOA uses $+e^2/r$ as in QM theories but $-e^2/r$ features in his two other bond approximations [8,9], see [10].

If it were not for the non-crossing rule at a critical separation $r_c$, ionic potentials would be wrong at $r>>r_0$ because of ionic dissociation limits. *Ionic asymptote* $H^+ + H^-$, with energy $IE_H + EA_H$ ($IE_H$ and $EA_H$ are ionization energy and electron affinity) differs from *atomic asymptote* $H+H$, with energy $2IE_H$ (this difference is 50 % if $EA \approx 0$). Only *electron transfer* can secure that $H+H$ goes over in charge-conjugated $H^+ + H^-$ or $H^- + H^+$. Although particle transfer can occur around $r_0$, if not $r_c$ due to the non-crossing rule, it remains highly unlikely at $r>>r_0$.

Particle transfers and disparate asymptotes are avoided with *intra-atomic charge inversion*, giving attractive $-e^2/r$ instead of repulsive $+e^2/r$ from first principles only [6]. Unfortunately, this model also has a price [6]: at an intermediary critical separation $r_0 \leq r_c \leq \infty$, a pair of neutral $H+H$ should go over in a pair of neutral $H+\underline{H}$, both pairs having comparable asymptotes $2IE_H$. However appealing for the asymptote problem and for the occurrence of attraction $-e^2/r$ instead of repulsion $+e^2/r$, chiral bonding mechanisms have other pro's and con's. One obstacle is Dirac pair annihilation, when extended to a pair of neutral H (*hydrogen*) and $\underline{H}$ (*antihydrogen*). While the uncertainty with respect to CPT remains for atomic H and $\underline{H}$, an advantage is that the so-called matter-antimatter asymmetry in the Universe would not longer be necessary [6,10].

Whatever the outcome of this debate, a search for theories, more refined than conventional ones, not in line with observed intra-molecular electron distributions [1-2,11], remains justified [4,6].

[1] K. Kreidi et al., Phys. Rev. Lett. **100**, 133005 (2008)
[2] J.D. Dunitz and P. Seiler, J. Amer. Chem. Soc. **105**, 7056 (1983)
[3] W.H.E. Schwarz, P. Valtazanos and K. Ruedenberg, Theor. Chim. Acta **68**, 471 (1985); W.H.E. Schwarz et al., Int. J. Quant. Chem. **29**, 909 (1986); K.L. Kunze and M.B. Hall, J. Am. Chem. Soc. **108**, 5122 (1986)
[4] G. Van Hooydonk, Phys. Rev. Lett. **100**, 159301 (2008)
[5] S. Weinbaum, J. Chem. Phys. **1**, 593 (1933)
[6] G. Van Hooydonk, Eur. Phys. J. D, **32**, 299 (2005)
[7] M. Born and R. Oppenheimer, Ann. Physik **84**, 457 (1927)
[8] M. Born and A. Landé, Verh. Deut. Phys. Ges. **20**, 210 (1918); M. Born, Ann. Physik **61**, 87 (1920).
[9] M. Born and J.E. Mayer, Z. Phys. **75**, 1 (1932)
[10] G. Van Hooydonk, arxiv:0803.2445
[11] D. Akoury et al., Science **318**, 949 (2007); F. Martin et al., Science **315**, 629 (2007)